\documentclass[letterpaper]{article} %
\usepackage{aaai25}  %
\usepackage{times}  %

\usepackage{makecell}
\usepackage[utf8]{inputenc}
\usepackage{graphicx}
\usepackage{comment}
\usepackage{fancyvrb}
\usepackage{xcolor}
\usepackage{subcaption}
\usepackage{xspace}
\usepackage{todonotes}
\usepackage{titling}
\usepackage{caption}
\usepackage{tabularx}

\newcommand{\url}[1]{{#1}}

\title{Interpreting Agent Behaviors in Reinforcement-Learning-Based Cyber-Battle Simulation Platforms}
\author{Jared Claypoole (SRI), Steven Cheung (SRI), Ashish Gehani (SRI),  \\ Vinod Yegneswaran (SRI), and Ahmad Ridley (Laboratory for Advanced Cybersecurity Research)}
\date{}

\begin{document}
\maketitle
\thispagestyle{plain}
\pagestyle{plain}

\begin{abstract}
We analyze two open source deep reinforcement learning agents submitted to the CAGE Challenge 2 cyber defense challenge, where each competitor submitted an agent to defend a simulated network against each of several provided rules-based attack agents.  We demonstrate that one can gain interpretability of agent successes and failures by simplifying the complex state and action spaces and by tracking important events, shedding light on the fine-grained behavior of both the defense and attack agents in each experimental scenario.
By analyzing important events within an evaluation episode, we identify patterns in infiltration and clearing events that tell us how well the attacker and defender played their respective roles; for example, defenders were generally able to clear infiltrations within one or two timesteps of a host being exploited.  
By examining transitions in the environment’s state caused by the various possible actions, we determine which actions tended to be effective and which did not, showing that certain important actions are between 40\% and 99\% ineffective.
We examine how decoy services affect exploit success, concluding for instance that decoys block up to 94\% of exploits that would directly grant privileged access to a host.  Finally, we discuss the realism of the challenge and ways that the CAGE Challenge 4 has addressed some of our concerns.
\end{abstract}

\newcommand{\cageone}{{CC1}\xspace}
\newcommand{\cagetwo}{{CC2}\xspace}
\newcommand{\cagefour}{{CC4}\xspace}

\newcommand{\cyborg}{{CybORG}\xspace}
\newcommand{\eireland}{{EIReLaND}\xspace}
\newcommand{\cyberbattlesim}{{CyberBattleSim}\xspace}
\newcommand{\farland}{{FARLAND}\xspace}

\section{Introduction}
In the modern digital landscape, the threat of cyberattacks is
ever-present. Conventional security strategies often find themselves
lagging behind the continuously advancing techniques of hackers. To
counter this, cyberdefense simulation and emulation platforms offer a
solution by creating virtual settings that model real-world networks,
where defensive tools and techniques can undergo ``battle testing''
through simulated cyberattacks. These platforms help unearth new
vulnerabilities, reduce reaction times, and equip network security
engineers with the skills to manage actual cyber threats effectively,
thereby offering organizations a proactive advantage against cyber
attacks.

Reinforcement Learning (RL) platforms~\cite{cyborg,CyberBattleSim, eireland-acd-2023,Farland-2021} mark 
a noteworthy advance in cyberdefense
simulation.  Their premise is centered around
AI-based agents that refine their strategies through a process of
trial and error, constantly tailoring their approach to counter their
opponents (i.e., attackers or defensive systems), which could be humans or
another AI system. This approach fosters a simulation environment that
is more dynamic and potentially mirrors real-life, intelligent human
attackers more closely than conventional, scripted attacks do.

There are several noteworthy advantages to these RL-based agent environments. The AI-driven
attacker, by incessantly learning and adapting, could mirror the evolving
dynamics of actual cyber threats. This provides security teams with
the opportunity to gauge their defense mechanisms against a smarter
and less predictable adversary, fostering the development of stronger
defensive strategies. Another key advantage lies in the development of
dynamic response strategies facilitated by AI-driven
simulations. These strategies are capable of autonomously adapting to
the specifics of an attack, offering a potent defense against complex,
multi-vector threats. This adaptability ensures that defensive
measures are not just reactive but also predictive, anticipating and
neutralizing threats before they manifest.

AI-driven simulators thus play a critical role in narrowing the
offense-defense gap, empowering defenders with attacker-like
insights. This strategic advantage allows for the anticipation of
future threats and the development of pre-emptive defenses, ensuring
that cybersecurity measures are several steps ahead of
potential attackers.  Nonetheless, there are associated challenges that need be considered.
First, we need to be aware of differences between simulation
and real-world environments, with respect to the mix of systems,
exploitable vulnerabilities, and response mechanisms.  Another significant
challenge is the opacity in understanding the AI-based attacker and
defender's decision-making process, which can complicate the learning
experience from simulations. Moreover, the real-world applicability of
these simulations could be limited, as attackers might have
different motivations and objectives, affecting the direct
transferability of simulation effectiveness to actual scenarios.

In this paper, we conduct a deep exploration of two top performing
RL-based cyberdefense agents that were produced in response to
the CAGE Challenge 2~\cite{cage_challenge_2} on the \cyborg simulation environment.  
We evaluate the choices made by the two defensive RL models with respect
to the two rules-based attack strategies implemented in the challenge.  
We then discuss limitations of this environment and potential opportunities
for improvement, including some that have been addressed in the latest version
of the CAGE Challenge 4~\cite{cage_challenge_4}~\footnote{CAGE Challenge 3 pertains to a scenario focusing on developing decentralized multi-agent defense for a drone-based environment that is quite different from those of CAGE 2 and CAGE 4.}.

\section{Background}
\begin{figure}
    \vspace{-0.2in}
    \centering
    \includegraphics[width=0.5\textwidth]{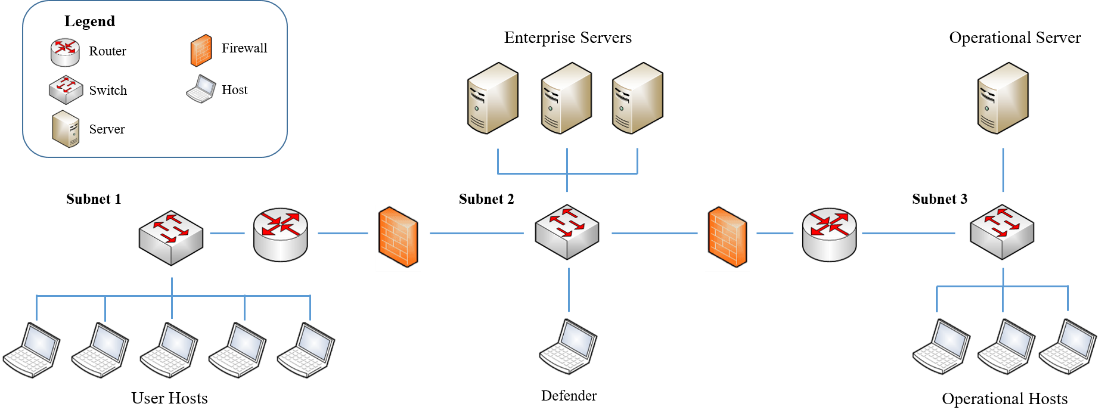}
    \vspace{-0.2in}
    \caption{
        Network diagram for Cage Challenge 2.
        Reproduced from Figure 1 of the \cagetwo announcement~\cite{cage_challenge_2}.
    }
     \vspace{-0.15in}
    \label{fig:network}
\end{figure}

\begin{table*}[t!]
\centering
\scriptsize
\begin{tabularx}{0.95\textwidth}{X|X}
    {\bf Blue Agent Actions} & {\bf Red Agent Actions}
    \\ \hline
    \textbf{Sleep} (perform no action).
    &   \textbf{Sleep} (perform no action).
    \\
    \textbf{Analyze} the detailed state of a given host.
    &   \textbf{DiscoverRemoteSystems}, revealing IP addresses of the hosts on a given subnet.
    \\
    Deploy one of seven possible \textbf{Decoy} services on a given host.
    &   \textbf{DiscoverNetworkServices}, revealing the services running on a given host.
    \\
    \makecell{
        \textbf{Remove} the red agent with non-privileged access from a given host.
        \\
        \textbf{Restore} a given host to a known state, removing any red agent while also\\ disrupting operation on that host.  This incurs a negative reward.
    }
    &  \makecell{
            Attempt to \textbf{Exploit} a service to gain access to a host (usually user-level access;\\ sometimes privileged-level access).\\
            \textbf{Escalate} privilege on a given host.\\ 
            \textbf{Impact} the critical Operational Server.
        } \\
\end{tabularx}
\caption{Summary of blue and red agent actions}
\label{tab:actions}
\end{table*}

\begin{figure}
    \centering
    \includegraphics[width=0.49\textwidth]{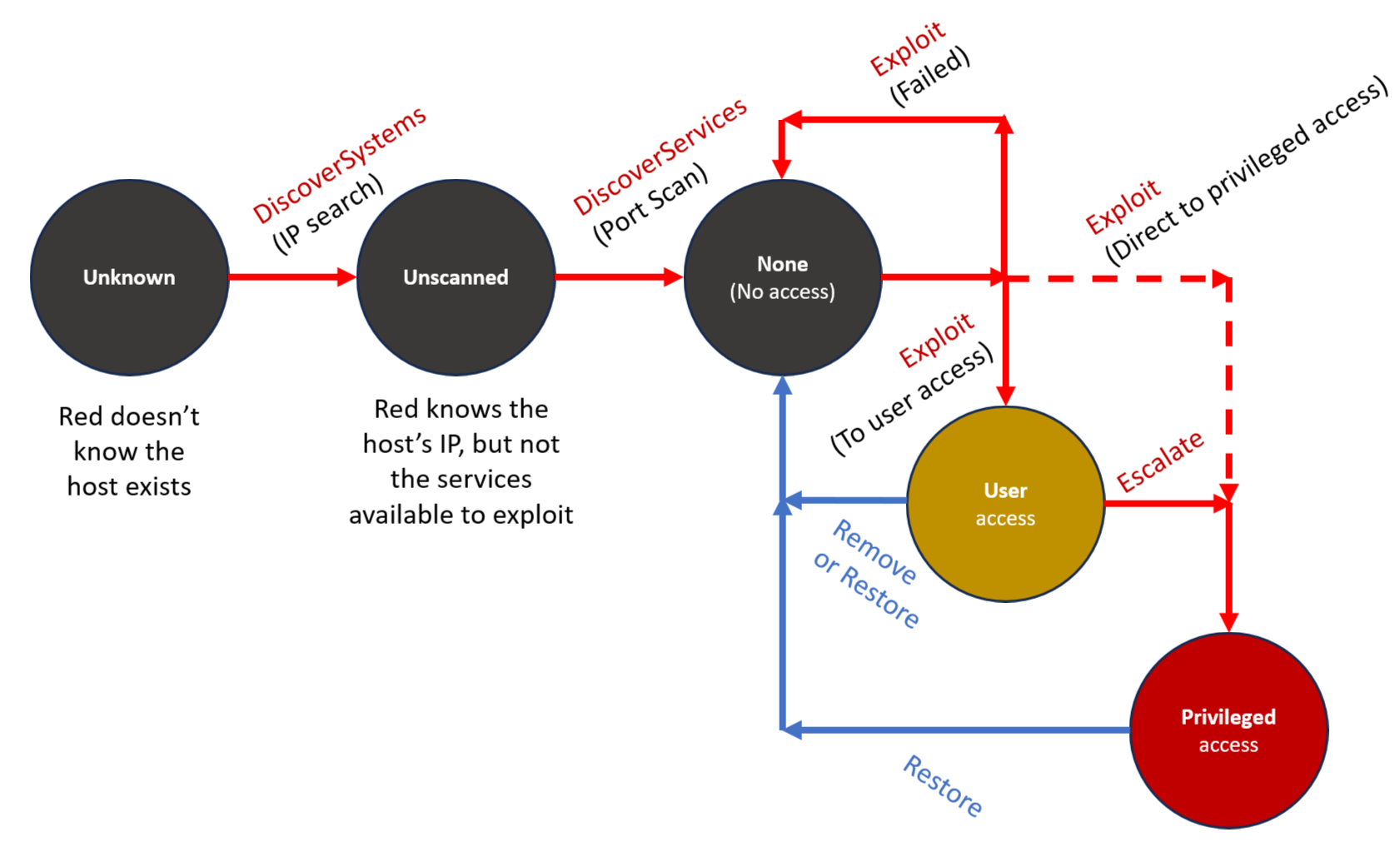}
     \vspace{-0.2in}
    \caption{
        State machine for each host in Cage Challenge 2.
        This does not include decoys,
        nor actions which don't change the state,
        such as escalate on a privileged host or blue's analyze action.
        Adapted from Figure 2 of the CC2 announcement~\cite{cage_challenge_2}.
    }
     \vspace{-0.2in}
    \label{fig:state-machine}
\end{figure}

The goal for Cage Challenge 2 (\cagetwo) competitors was to submit a blue agent which could successfully defend the simulated network (see Figure~\ref{fig:network}) against a rules-based red agent.
The challenge is naturally posed as a reinforcement learning problem for a blue agent:
at each time step, the agent first receives an observation of its environment, next chooses an action to perform, and finally receives a reward, before moving on to the next time step.
In \cagetwo, the blue agent receives negative reward every turn for each host to which the red agent has privileged access, according to how valuable the host is considered to network operations.
There is also exactly one blue action, restore, that incurs a negative reward.
Table~\ref{tab:actions} provides a summary of actions for both agents and
Figure~\ref{fig:state-machine} provides a visualization of the different possible states for each host and how the red and blue actions affect the state.  The blue agents are evaluated against three distinct rules-based red agents.  The first is the Sleep agent, which simply performs the sleep action at each time step.  Next is the B-line agent, which has prior knowledge of the network structure, allowing it to infiltrate more effectively.  Finally, the Meander agent has no knowledge of the network structure and thus must search each subnet for a host that points to an IP address on another subnet.

\section{Evaluation Results}
We analyze the two open source top performing submissions to the challenge: team Cardiffuni~\cite{cardiffuni_submission}, which scored first place in the challenge, and team Mindrake~\cite{mindrake_submission}, which scored third in the challenge.
We use these submitted agents to investigate the following research questions:
\begin{itemize}
    \item RQ1 -- What techniques can we broadly use to analyze agent behavior in \cagetwo?
    \item RQ2 -- What conclusions can we make from examining the behavior of individual episodes?
    \item RQ3 -- What conclusions can we make from the official \cagetwo evaluations?
    \item RQ4 -- What conclusions can we make from examining state transitions among hosts as a result of the agents' actions?
    \item RQ5 -- What conclusions can we make from examining infiltration and clearing events?
    \item RQ6 -- What conclusions can we make from examining the effect of decoy services on attempted infiltrations?
    
\end{itemize}

\subsection{Analysis Methods}

\begin{figure*}[t!]
    \centering
    \includegraphics[width=0.8\textwidth, trim={0 0 0 1cm}, clip]{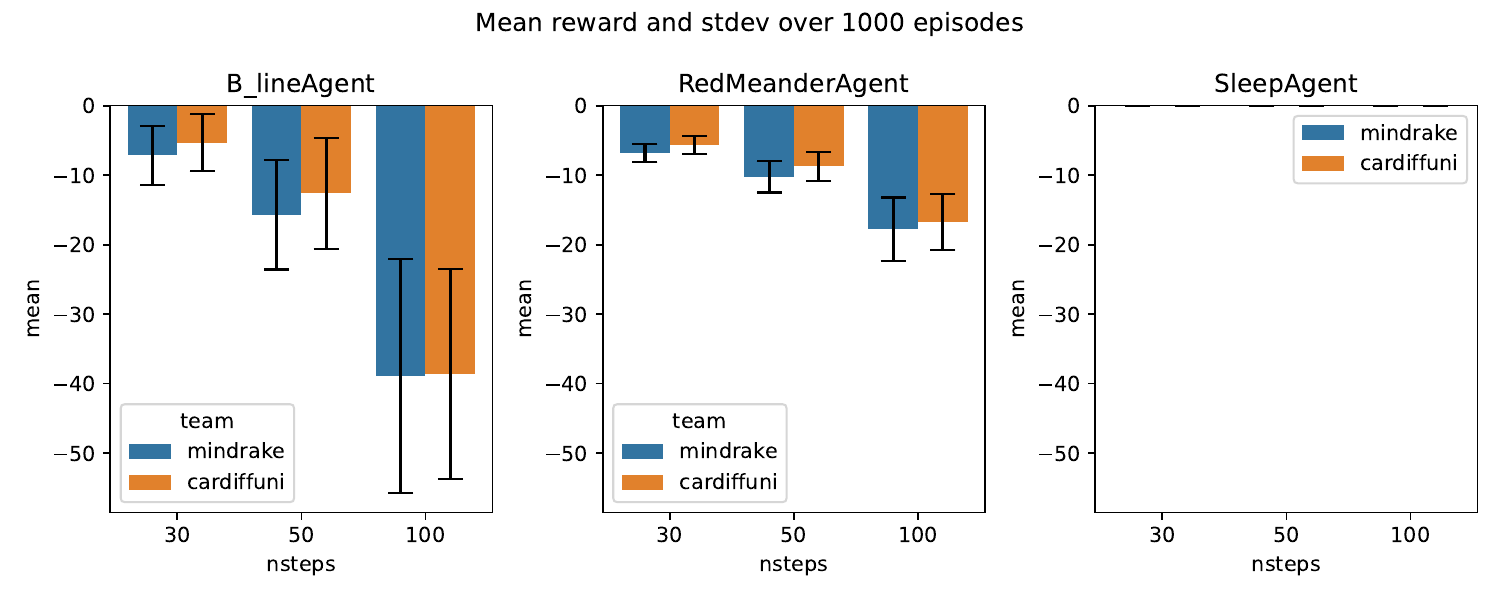}
    \vspace{-0.2in}
    \caption{
        Mean reward and standard deviation for each combination of three red agents and three different numbers of steps per episode.
        1000 episodes were run for each configuration.
        We see that Cardiffuni narrowly outperforms Mindrake against the B-line and Meander agents,
        and both consistently achieve the maximal reward of zero against the Sleep agent.
    }
    \label{fig:quant-eval-results}
    \includegraphics[width=0.8\textwidth]{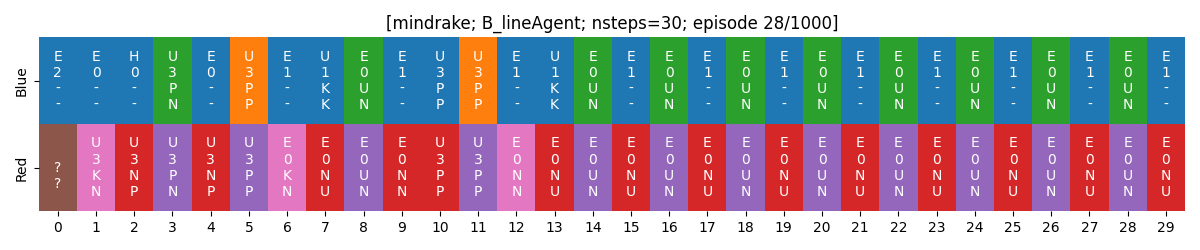}\\
    \begin{minipage}{0.17\textwidth}
    \includegraphics[width=0.75\textwidth]{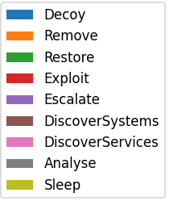}
    \end{minipage}
    \scriptsize
    \begin{tabular}{c||cc||cc}
    Text labels & Host group & Abbrev & Access state & Abbrev\\ \hline
    Host group & User & U & Undefined & ?\\
    Host index & Enterprise & E & Unknown & - \\
    Prev state & Defender & D & Known & K \\
    Post state & Op\_Host & H & None & N \\
    & Op\_Server & S & User & U \\
    & & & Privileged & P
    
    \end{tabular}
     \vspace{-0.1in}
    \caption{
        A visualization of a single episode
        and legends explaining the colors (actions taken) and text labels in the visualization.
        The cells on top represent blue agent actions, and the cells below represent the red agent actions which follow.  The horizontal axis is time steps.
        Most striking in the visualization of this episode (as with most others) is the pattern of back and forth behavior in the second half of the episode, where the red agent exploits a host and the blue agent immediately responds with a restore action.
    }
    \label{fig:single-ep-viz}
     \vspace{-0.2in}
\end{figure*}

\textbf{Analysis using the official evaluation script.}
First we use the evaluation script provided by the challenge authors, which each team modified slightly to load their agent.
This script ran the environment in nine different run configurations -- three different red agents (Sleep agent, B-Line agent, and Meander agent) and three different episode lengths (30, 50, and 100 time steps per episode).  Each of the nine configurations was run 1000 times.
This script outputs a text file containing the actions the blue and red agents performed at each time step throughout each episode, and the total reward at the end of the episode.
The output file also contains headings describing each of the nine run configurations, and the mean reward for each configuration, as plotted in Figure~\ref{fig:quant-eval-results}.
The teams were given a final score equal to the sum of these nine mean rewards.
Our analysis agrees qualitatively with the official CAGE evaluations, but we were not able to reproduce their quantitative results.

We use the evaluation script's output to aggregate the actions performed by each agent.
We can also use the sequence of actions to determine which actions each agent performed in response to the other's, but without information about the state of the system before and after these actions, it's difficult to determine how effective these actions are.

\textbf{Tracking the state of each host.}
As our second method of analysis, we examine the state of the system, making use of the fact that the full state factorizes into the product of states of the 12 exposed hosts.
We modify the evaluation script to record the state of each host in the simulation separately, primarily using what the \cyborg environment calls the ``true table'' -- a table describing the ground truth state of each of the 12 hosts.
We obtain this true table before every blue action
(where at each time step, first the blue agent performs an action, followed by the red agent).
Then, every time the state of the host changes (including initialization) we record that state in our table of state changes.
We are then able to associate each red and blue action with the state of the relevant host before and after the time step when that action was performed.
(These are the same state if the action didn't change the state of a host, or if the other agent performed an action which interfered with the state so as to cancel out the effect of the action.)
Note that some actions do not act on a single host.
These include the sleep action, which can largely be ignored, and red's DiscoverRemoteSystems action, which acts on a subnet rather than a single host, revealing all hosts connected to that subnet.
The DiscoverRemoteServices action actually does change the state of the hosts on the subnet it acts on -- these hosts are no longer unknown to the red agent -- but not in a way that affects the actions of the blue agent.

\textbf{Examining individual episodes.}
We produce visualizations, such as Figure~\ref{fig:single-ep-viz}, representing the alternating blue and red agent actions within a single episode.  The color represents the action taken, while the annotations in each box represent the subnet and host upon which the agent is acting, as well as that host's state before and after the current time step.

\textbf{Examining state transitions.}
Given tracked state data,
we can then examine the state transitions caused by actions.  Because we only obtain the ground truth state at the start and end of every time step, in general we can't say whether it was the action performed by red, blue, or a combination of the two that caused a given change in state.  However, when the two agents act on different hosts within a given time step, the actions are independent\footnote{
    Again, this isn't quite true for red's DiscoverRemoteSystems action, as it acts on a subnet rather than a single host, but this action is easy to take into account, as it (1) has a clear effect and (2) doesn't interfere with the blue agent's actions.
}, and it is clear what effect each action had on the corresponding host.

The data we have to work with for a state transition are two actions (one for each red and blue), a state at the start of the time step, before the blue agent acts, and a state at the end of the time step, after both agents have acted.
In cases where the agents act on different hosts within the same time step,
we can focus on the two agents' actions independently.

\textbf{Examining infiltrations.}
The red agent is constantly attempting to infiltrate hosts,
and the blue agent is correspondingly trying to mitigate such infiltrations.
We track these infiltration events: when they begin and end, how long they last, and on which subnets they occur.

\textbf{Examining Decoys.}
Finally, we track Decoy services deployed by the blue agent,
including how many of each type of decoy (and the total number of decoys) were deployed at the time of each attempted exploit.

\subsection{Results from single episode visualizations}

We observed the following qualitative behaviors from visualizations shown in Figure~\ref{fig:single-ep-viz}:
\begin{itemize}
    \item When the red agent infiltrates a host on the Enterprise subnet, the blue agent immediately responds with a restore action, which is often an overkill response to a user-level infiltration.
    \item When not responding to a successful exploit, the blue agent typically performs its most prevalent action.  For Mindrake this is decoy, and for Cardiffuni this is remove.
    \item After gaining a privileged foothold on the User subnet, the red agent often repeatedly performs redundant exploits on the already infiltrated host.
\end{itemize}

\subsection{General results from official evaluation output}
Team Cardiffuni narrowly wins over team Mindrake in all six non-sleep run configurations (and the two teams tie against the sleep agent) as shown in Figure~\ref{fig:quant-eval-results}.
Meanwhile both teams achieve exactly zero reward against the sleep agent, which is the best possible scenario.  While they could have performed any action except Restore to achieve this, in practice the blue agents also performed the sleep action at nearly every time step.

\begin{figure}[t!]
    \centering
    \includegraphics[width=0.4\textwidth]{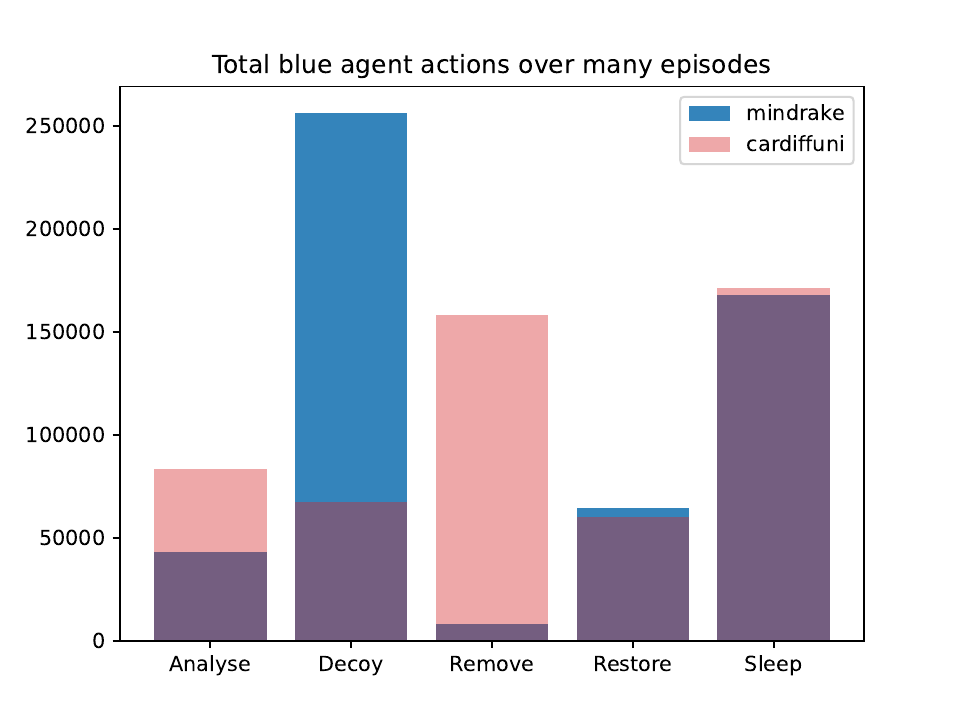}
     \vspace{-0.2in}
    \caption{
        Counts of each type of action taken by each of the two blue agents
        throughout the 9000 evaluation episodes.
    }
    \label{fig:action-counts-blue}

    \hspace{\floatsep}
    \centering
    \includegraphics[width=0.4\textwidth]{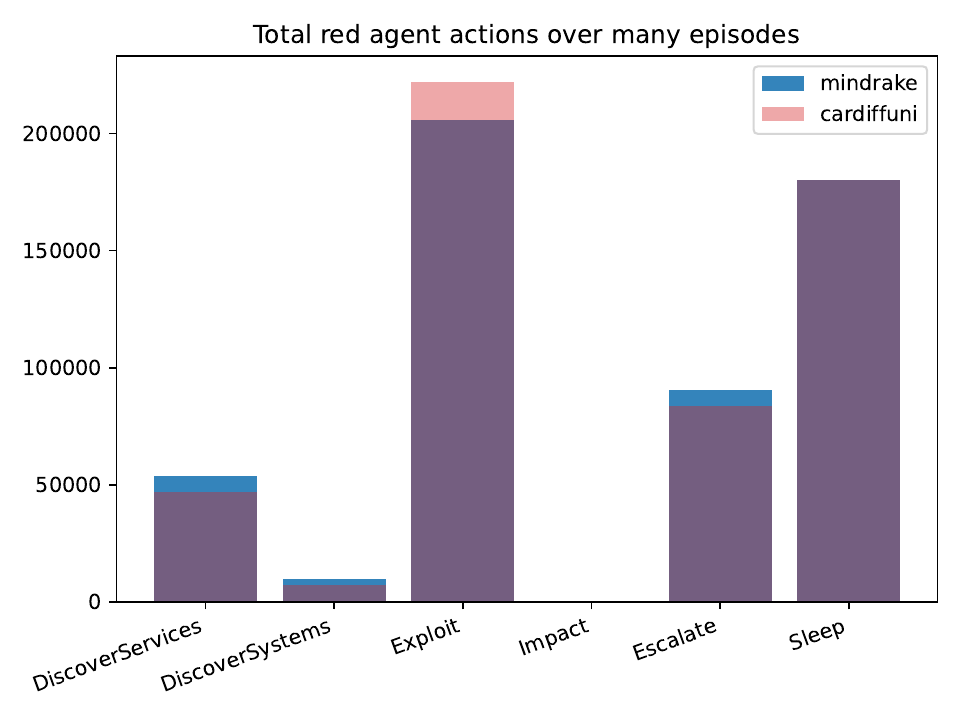}
     \vspace{-0.2in}
    \caption{
        Counts of each type of action taken by all three red agents (aggregated)
        throughout the 9000 evaluation episodes.
        Red agents behaved similarly despite differing strategies from the two teams.
    }
    \label{fig:action-counts-red}
     \vspace{-0.2in}
\end{figure}

Figure~\ref{fig:action-counts-blue} shows that Team Mindrake relies on a strategy of deploying Decoy services, while Team Cardiffuni relies more heavily on Analyzing hosts and attempting to Remove the red agent from hosts.  While at first it appears that the blue agents often take Sleep actions, in practice they only ever perform these actions against the red sleep agent. Figure~\ref{fig:action-counts-red} shows that despite the differences in strategy for the two teams, the red agents act and affect the environment similarly when opposing each team.

\subsection{State transition analysis results}
Based on the state transition statistics plotted in Figure~\ref{fig:tgrid-exploit}-e,
we observe the effectiveness of various agent actions.

Among successful exploits, the ratio of User to Privileged resulting access is roughly 3 or 4 to 1 as shown in Figure~\ref{fig:tgrid-exploit}.
We can see that roughly 10-15\% of exploit actions are taken against privileged hosts, which would only be effective if the blue agent removed/restored the host immediately before.  Remove actions are very rarely effective.
As illustrated in Figure~\ref{fig:tgrid-remove},
only 0.12\% of Mindrake's and 0.26\% of Cardiffuni's actions are removes which act on a host with User-level access.
This is especially devastating to Cardiffuni, as roughly 56\% of their actions are Removes.

The majority (roughly 65-70\%) of the time, Restore actions act on hosts with User level access, meaning Remove would have been just as effective without incurring the negative reward that Restore does, as shown in Figure~\ref{fig:tgrid-remove}. The red agent is inefficient with the DiscoverServices action: according to Figure~\ref{fig:tgrid-services}, roughly 40\% act on previously scanned hosts. Roughly 40-45\% of Escalate actions act on states that are already privileged. This is never an effective action, as shown in Figure~\ref{fig:tgrid-escalate}.

\newcommand{\transitionCaptionSize}{\small }
\begin{figure*}[t!]
    \centering
    \begin{subfigure}{0.32\textwidth}
        \includegraphics[width=0.99\textwidth]{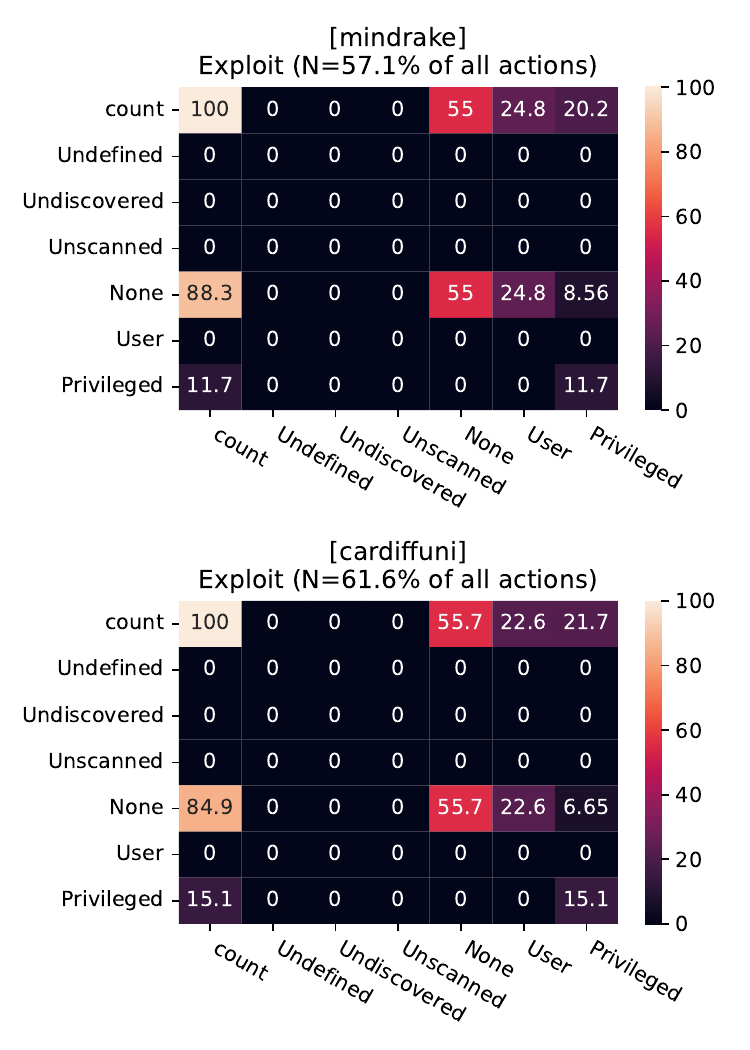}
        \caption{\transitionCaptionSize{}Exploit}
        \label{fig:tgrid-exploit}
    \end{subfigure}
    \begin{subfigure}{0.32\textwidth}
        \includegraphics[width=0.99\textwidth]{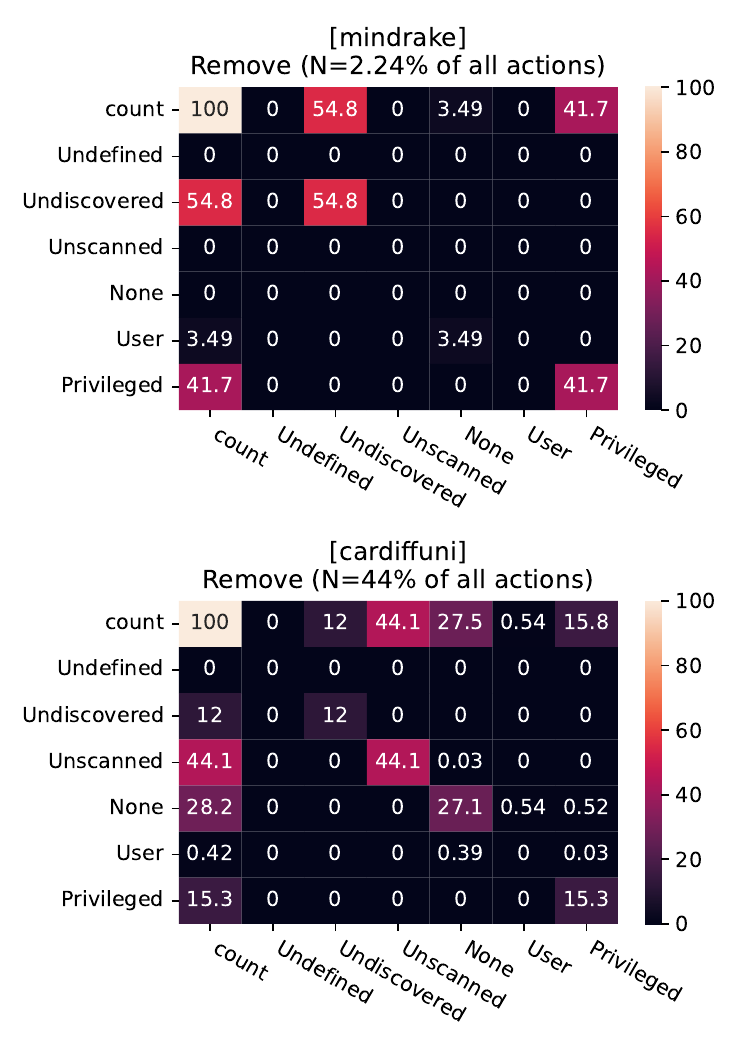}
        \caption{\transitionCaptionSize{}Remove}
        \label{fig:tgrid-remove}
    \end{subfigure}
    \begin{subfigure}{0.32\textwidth}
        \includegraphics[width=0.99\textwidth]{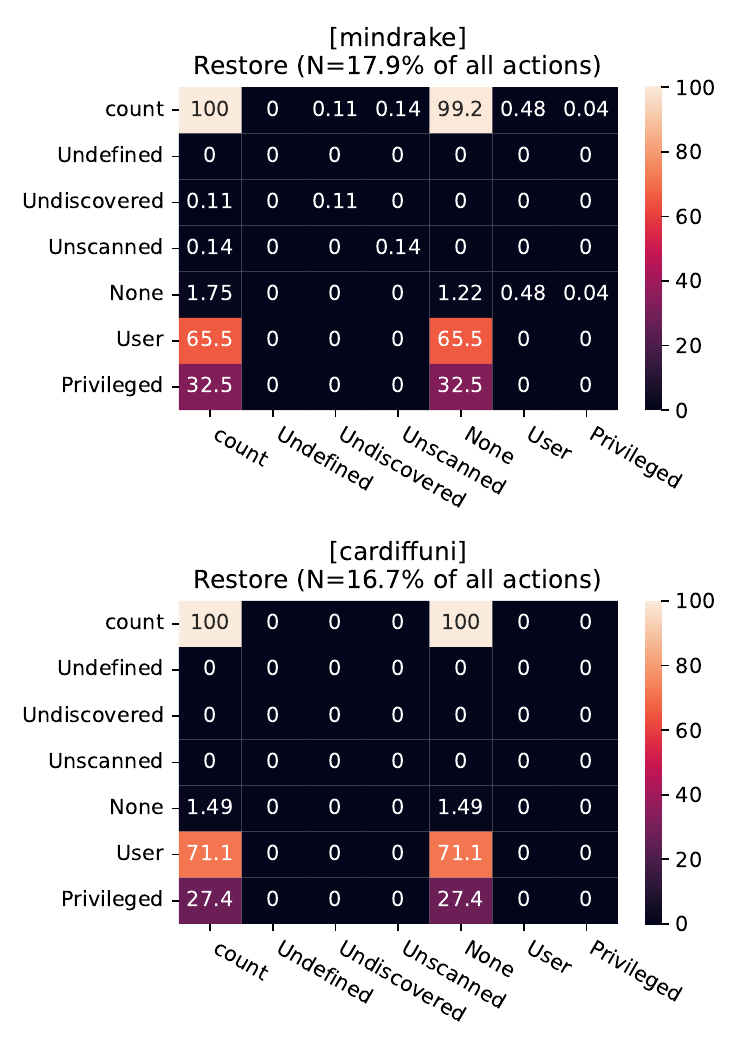}
        \caption{\transitionCaptionSize{}Restore}
        \label{fig:tgrid-restore}
    \end{subfigure}
    \begin{subfigure}{0.32\textwidth}
        \includegraphics[width=0.99\textwidth]{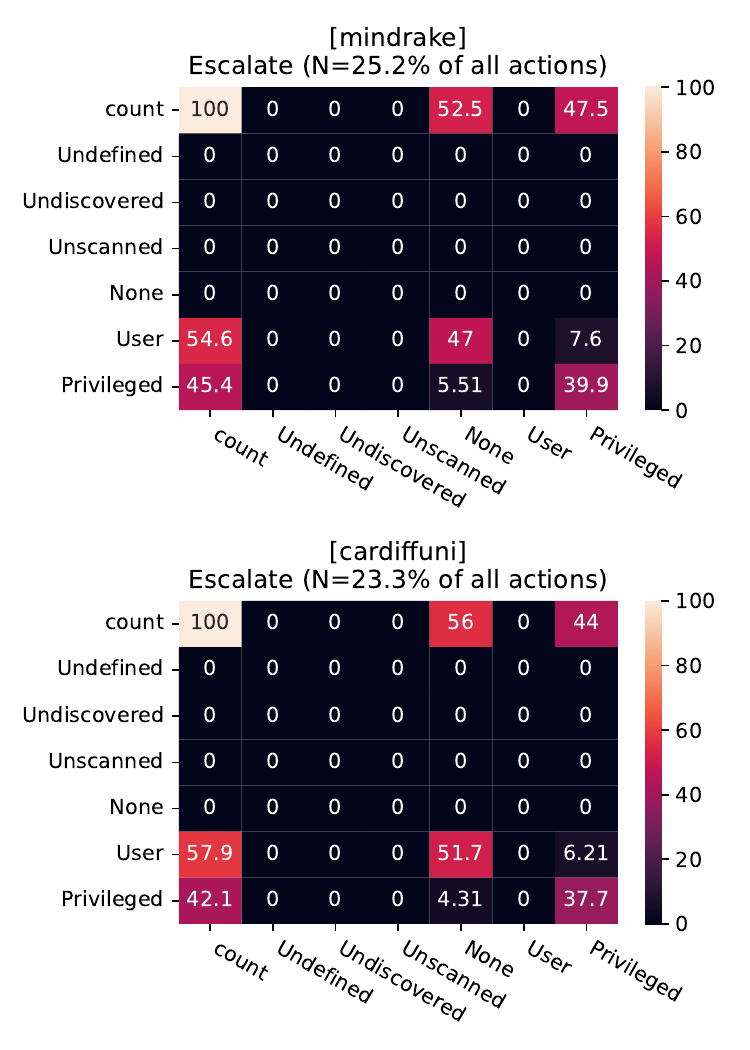}
        \caption{\transitionCaptionSize{}Escalate}
        \label{fig:tgrid-escalate}
    \end{subfigure}
    \begin{subfigure}{0.32\textwidth}
        \includegraphics[width=0.99\textwidth]{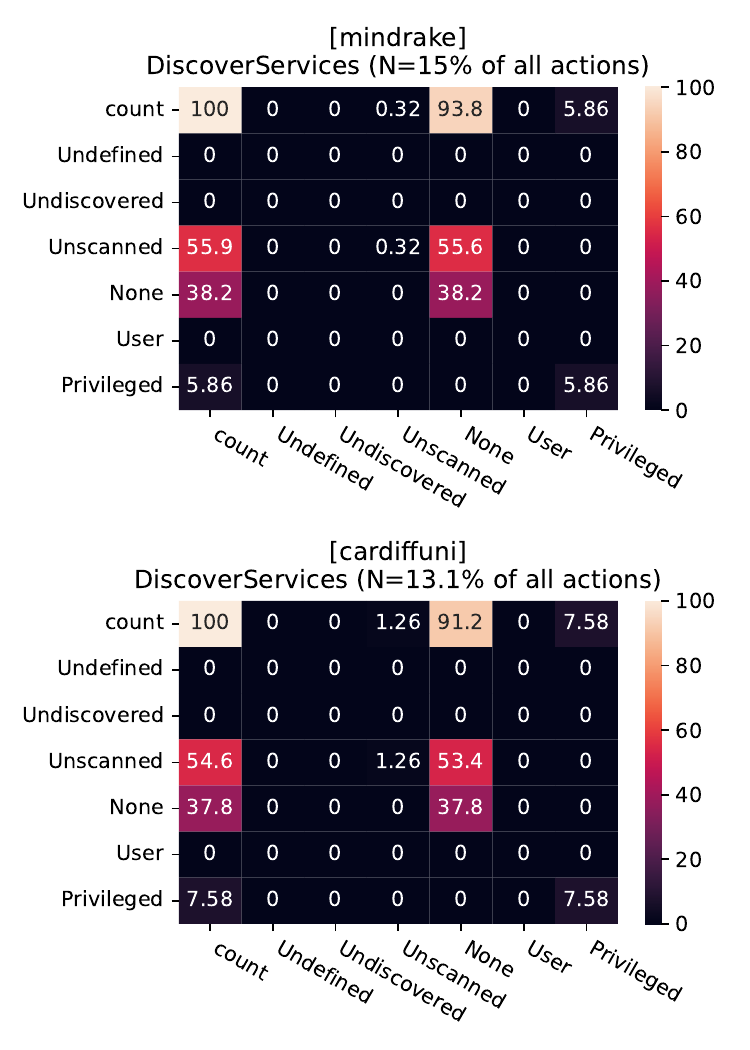}
        \caption{\transitionCaptionSize{}DiscoverServices}
        \label{fig:tgrid-services}
    \end{subfigure}
    \label{fig:tgrid-all}
     \vspace{-0.1in}
    
    \caption{
        Individual host state transitions for various red or blue agent actions.
        Prior states are listed on the left and post states on the bottom.
        Rows/columns labeled ``count'' indicate the percentage of all prior/post states which correspond to the given row/column;
        these are equal to the sum of the corresponding row/column.
        The remaining cells indicate the percentage of all actions of the type indicated in the plot's title that resulted in the corresponding state transition.
        Note: When the titles refer to the percentage of all actions, they describe the percentage of all actions in episodes that do not involve the red sleep agent, because such episodes are mostly trivial.
    }
     \vspace{-0.2in}
\end{figure*}

\subsection{Infiltration analysis results}
\begin{figure}[t!]
    \centering
    \includegraphics[width=0.49\textwidth, trim={0 0 0 1cm}, clip]{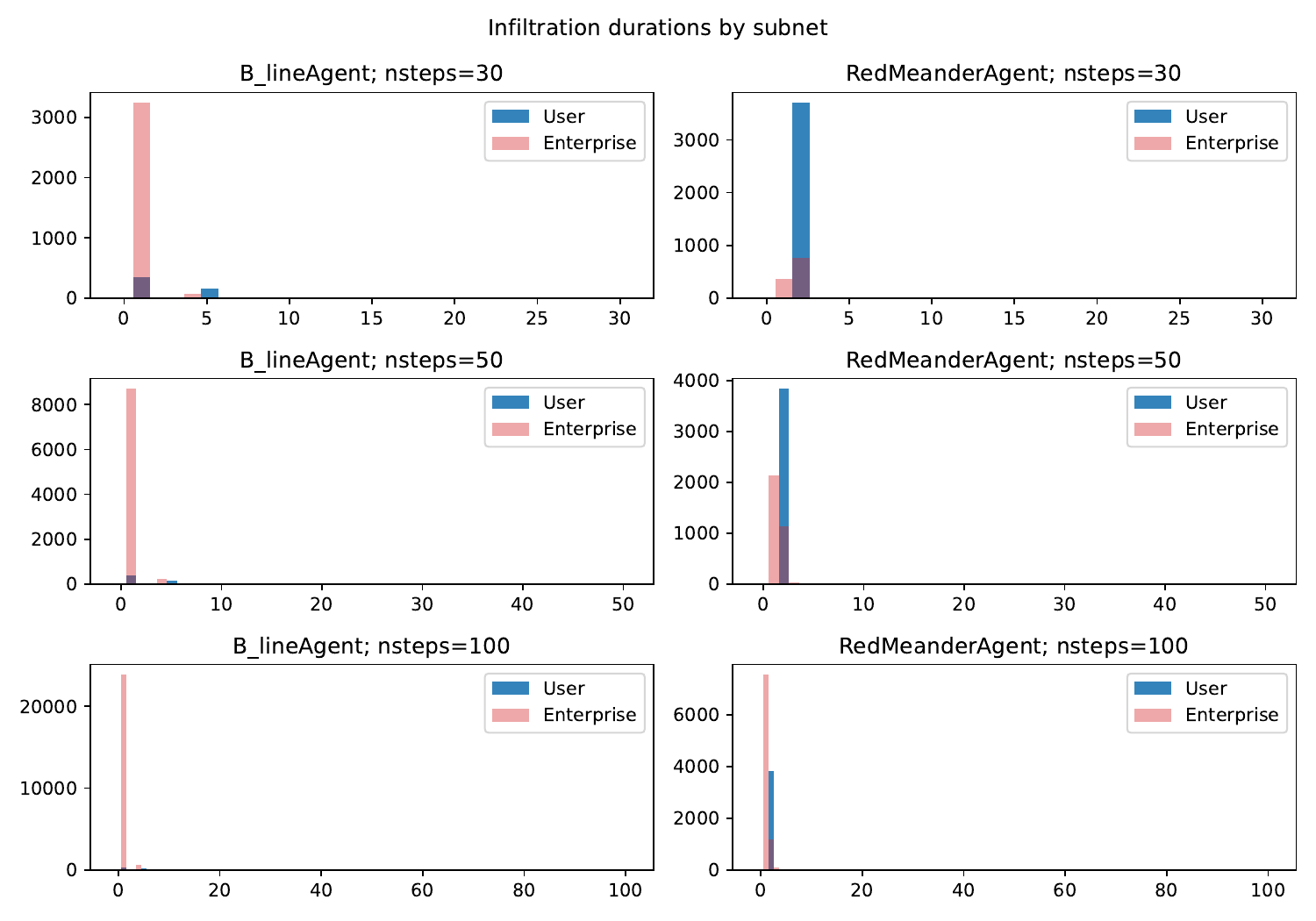}
     \vspace{-0.2in}
    \caption{
        Mindrake:  Infiltration durations on each subnet for each run configuration.
    }
    \label{fig:mindrake-infil-durations}
    \includegraphics[width=0.49\textwidth, trim={0 0 0 1cm}, clip]{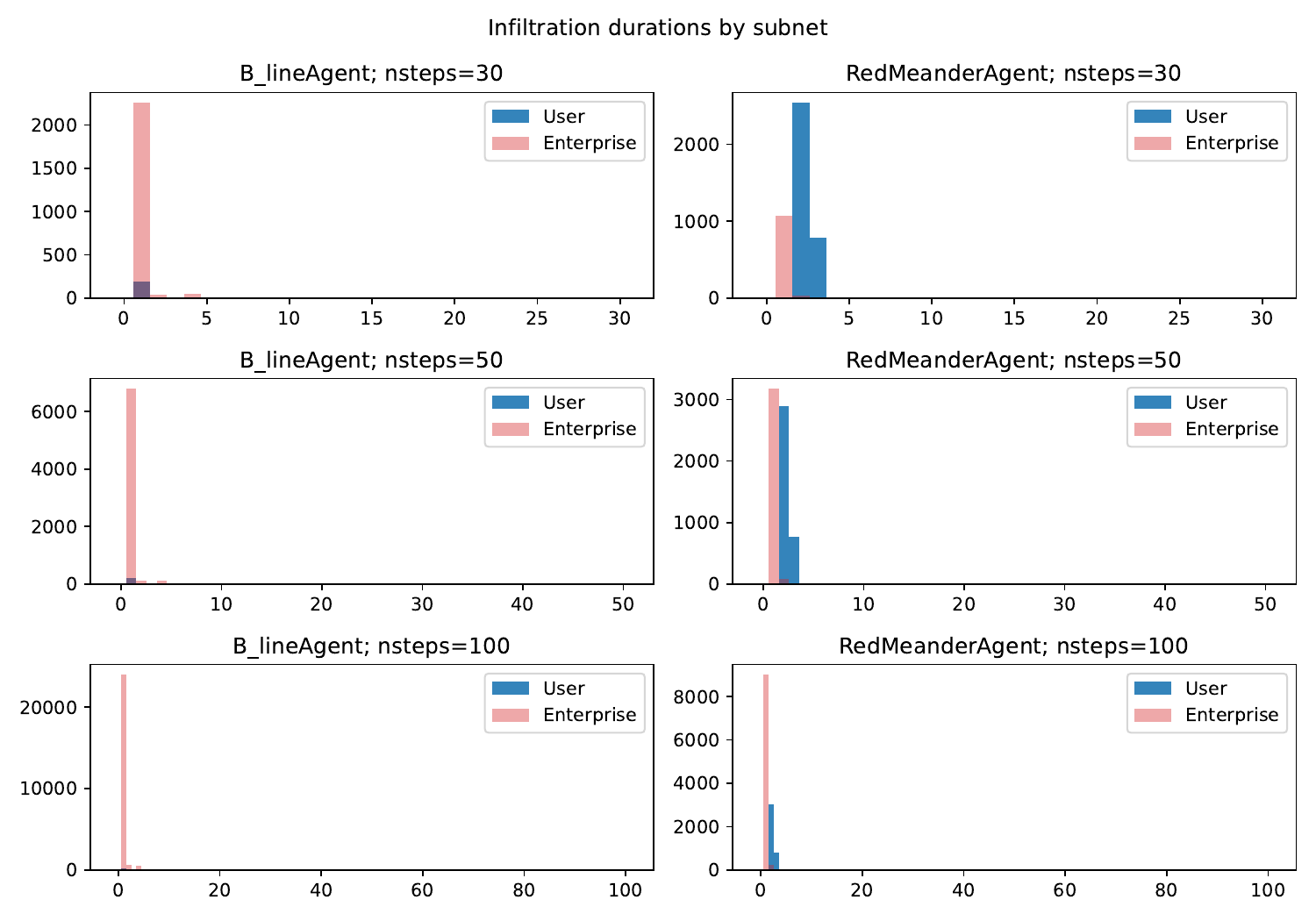}
     \vspace{-0.2in}
    \caption{
        Cardiffuni:  Infiltration durations on each subnet for each run configuration.
        We can see the results are qualitatively similar to Mindrake's.
    }
    \label{fig:cardiffuni-infil-durations}
\end{figure}
\begin{figure}[t!]
    \centering
    \vspace{-0.1in}
    \includegraphics[width=0.49\textwidth]{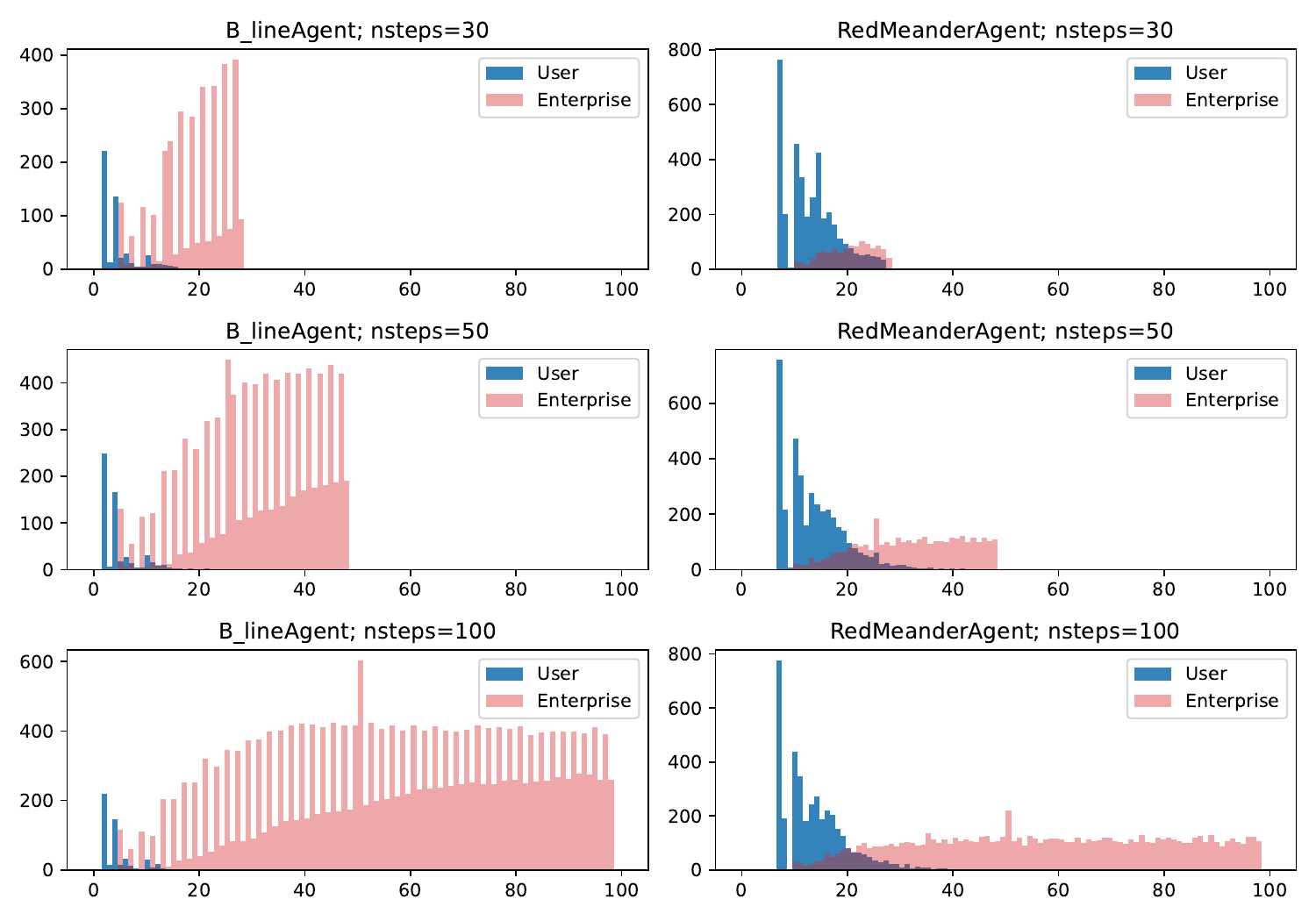}
     \vspace{-0.2in}
    \caption{
        Mindrake:  Infiltration starting steps on each subnet for each run configuration.
    }
    \label{fig:mindrake-infil-starts}
    \includegraphics[width=0.49\textwidth]{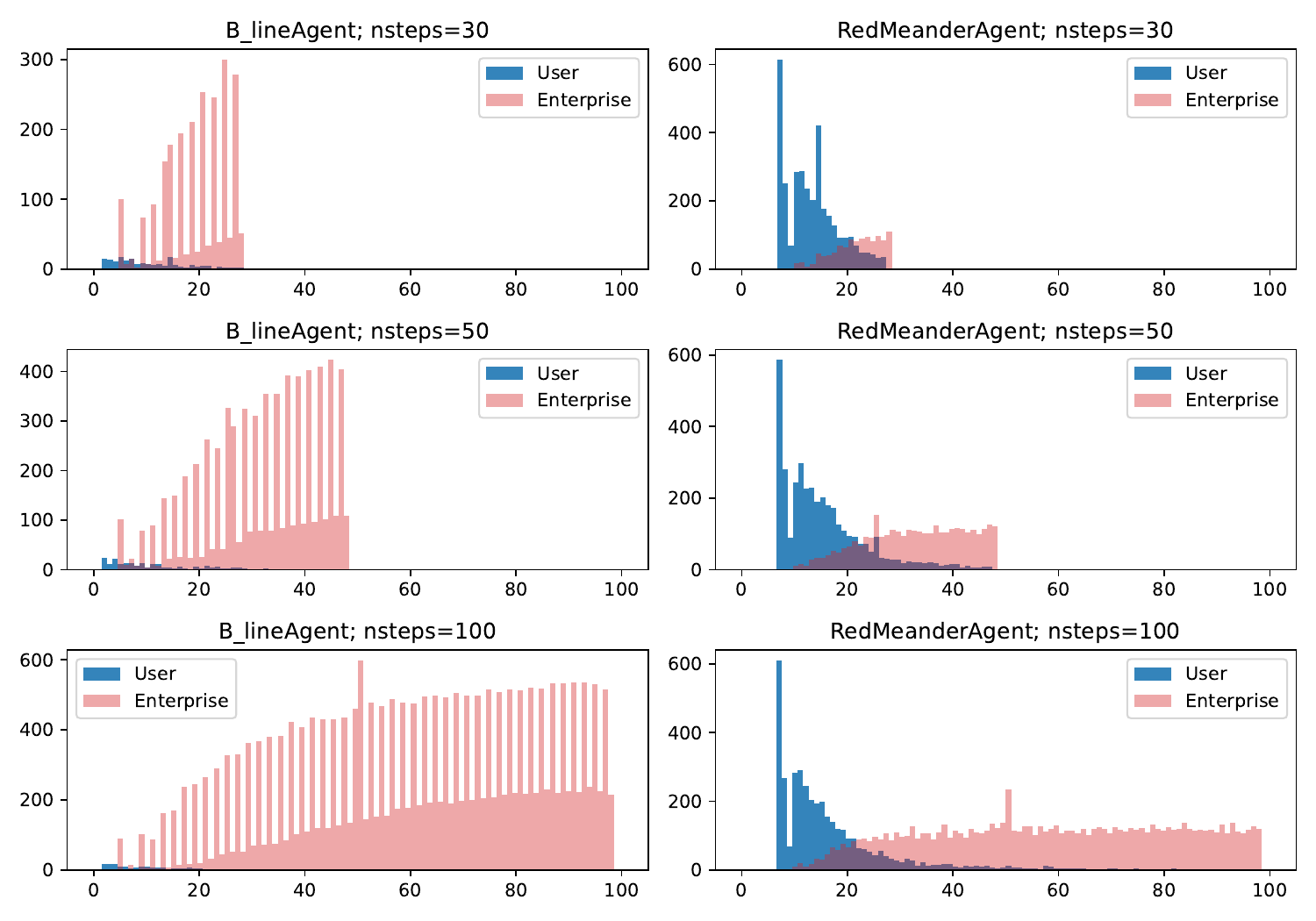}
     \vspace{-0.2in}
    \caption{
        Cardiffuni:  Infiltration starting steps on each subnet for each run configuration.
    }
    \label{fig:cardiffuni-infil-starts}
\end{figure}
Both blue agents are generally able to clear the red agent from the host within a few time steps
as shown in Figures~\ref{fig:mindrake-infil-durations} and \ref{fig:cardiffuni-infil-durations}.
The B-line agent focuses on the Enterprise subnet, rarely infiltrating the User subnet.
The Meander agent focuses on the User subnet for shorter episodes, but for longer episodes it focuses more and more on the Enterprise subnet.
Despite differences in the agents' strategies, their infiltration statistics were qualitatively similar.

Infiltrations on the User subnet occur toward the beginning of an episode; infiltrations on the Enterprise subnet begin later.
The red B-line agent is more quickly able to move from the User to the Enterprise subnet than is the Meander agent, as expected from the former's prior knowledge of the network structure.

\subsection{Decoy analysis results}

In the presence of decoys, successful privileged-level exploits sharply drop for both nontrival red agents against both teams as shown in Figure~\ref{fig:ndecoys-overall}.
Often this occurs at the cost of additional successful user-level exploits, meaning the blue agents are prioritizing blocking privileged-level exploits over user-level ones.

Perhaps a surprising feature of these figures is their non-monotonicity.
For team Mindrake, whose strategy relies much more on decoys, this non-monotonicity is almost entirely explained by splitting the plots by subnet.
Most exploits on the User subnet (Figure~\ref{fig:ndecoys-user}, left) hit zero decoys, whereas exploits on the Enterprise subnet never hit zero decoys (Figure~\ref{fig:ndecoys-enterprise}, left).
The differences between the B-line and Meander agents in these plots is explained by two features of the Meander agent's behavior:
(1) the Meander agent performs more exploits on the User subnet,
meaning it has many more attempted exploits at zero decoys; and
(2) by the time the Meander agent gets to the Enterprise subnet, there are typically two or more decoys (the peak is three), leading to the remaining non-monotonicity for the Meander agent's Enterprise subnet exploits.

For team Cardiffuni, the non-monotonicities in the figures are not well explained by splitting them by subnet.
However, we find these non-monotonicities are fairly well correlated with counts of the number of decoys present on each subnet at the end of each episode as shown in Figure~\ref{fig:cardiffuni-nhosts-ndecoys}.
This indicates the decoys are deployed on the two subnets in non-monotonic amounts.

\newcommand{\decoyfigwidth}{{0.45\textwidth}}
\begin{figure*}[t!]
    \centering
    \includegraphics[width=\decoyfigwidth]{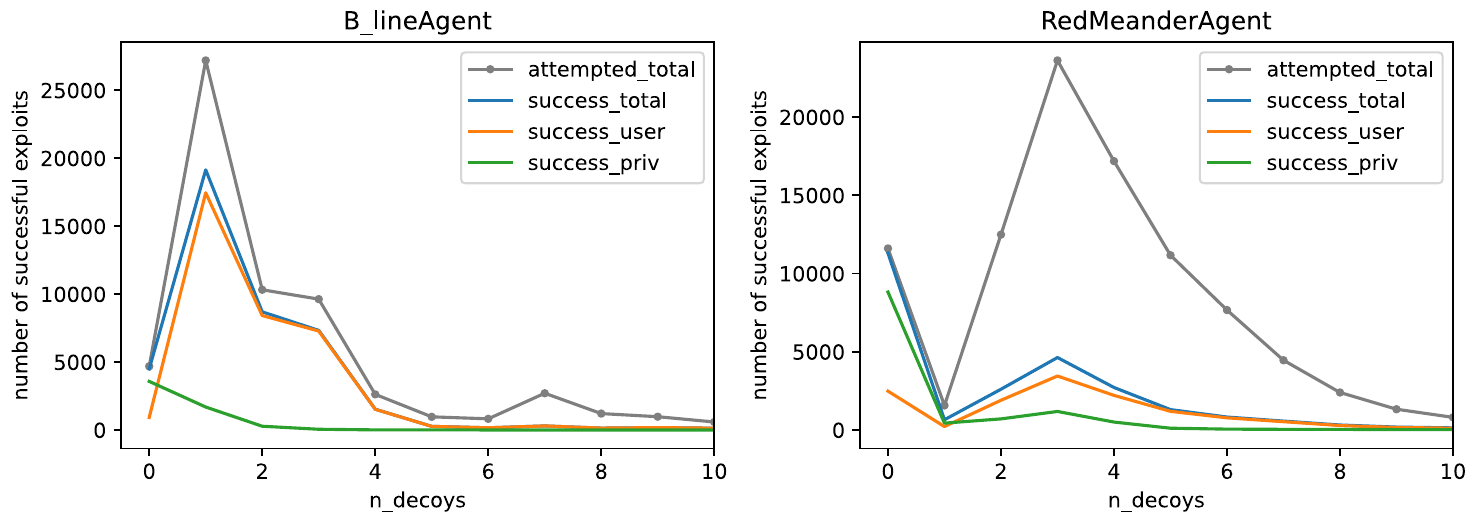}
    \hspace{1cm}
    \includegraphics[width=\decoyfigwidth]{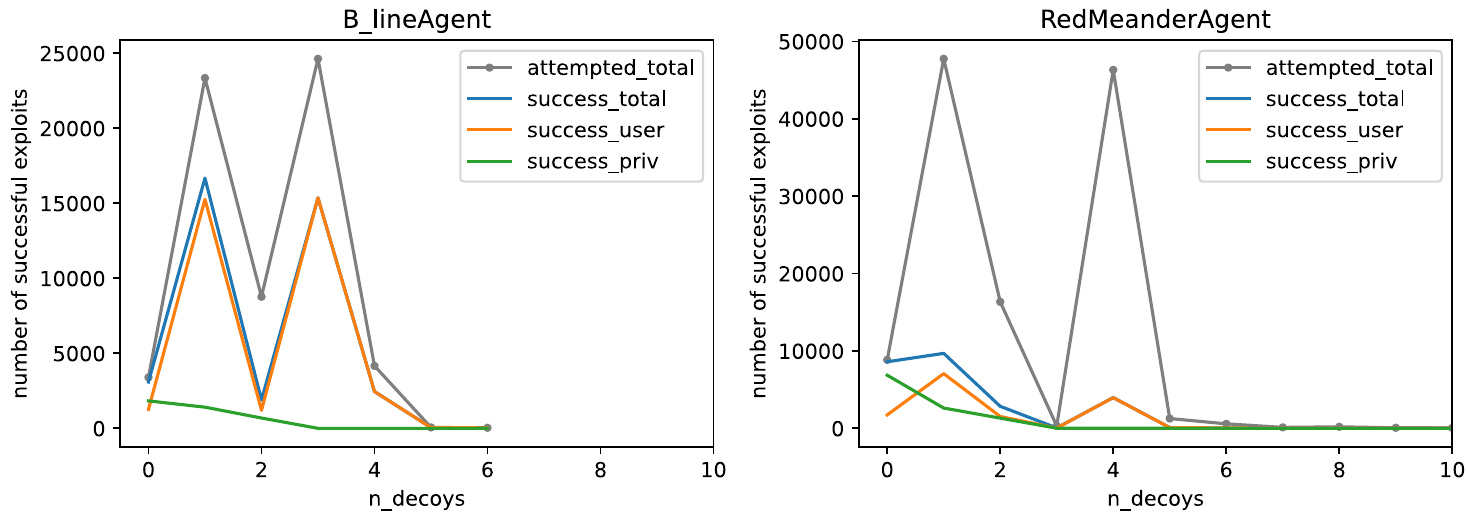}
     \vspace{-0.1in}
    \caption{(Mindrake, left; Cardiffuni, right):  Number of successful exploits vs number of decoys on the exploited host against each red agent.}
    \label{fig:ndecoys-overall}
    
    \includegraphics[width=\decoyfigwidth]{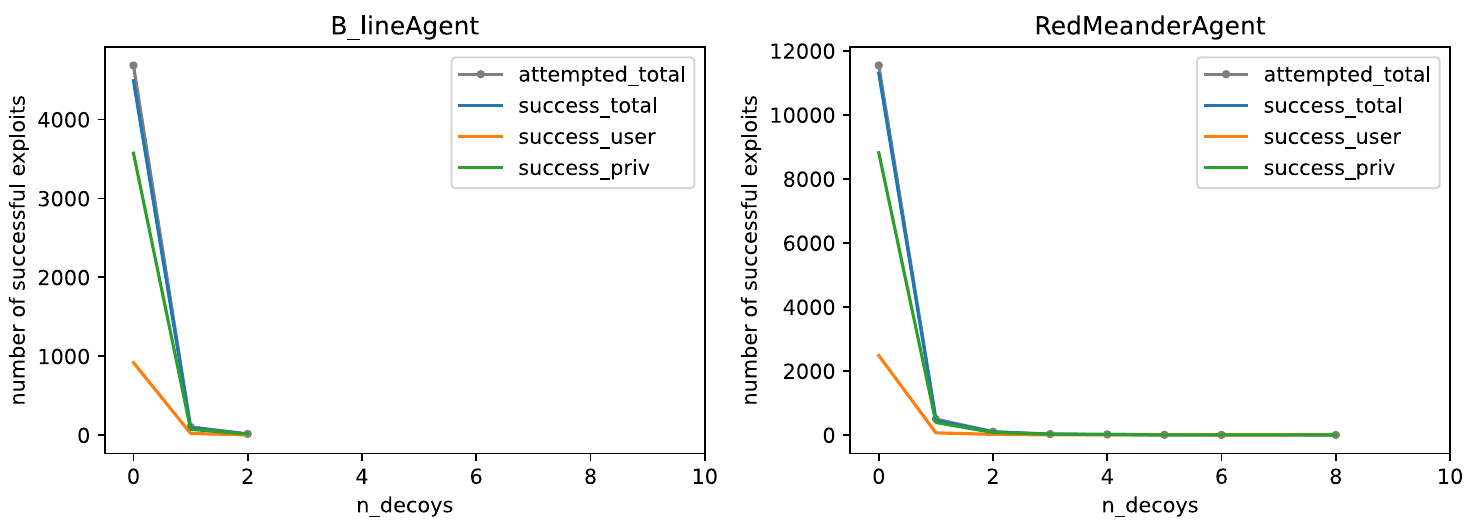}
    \hspace{1cm}
    \includegraphics[width=\decoyfigwidth]{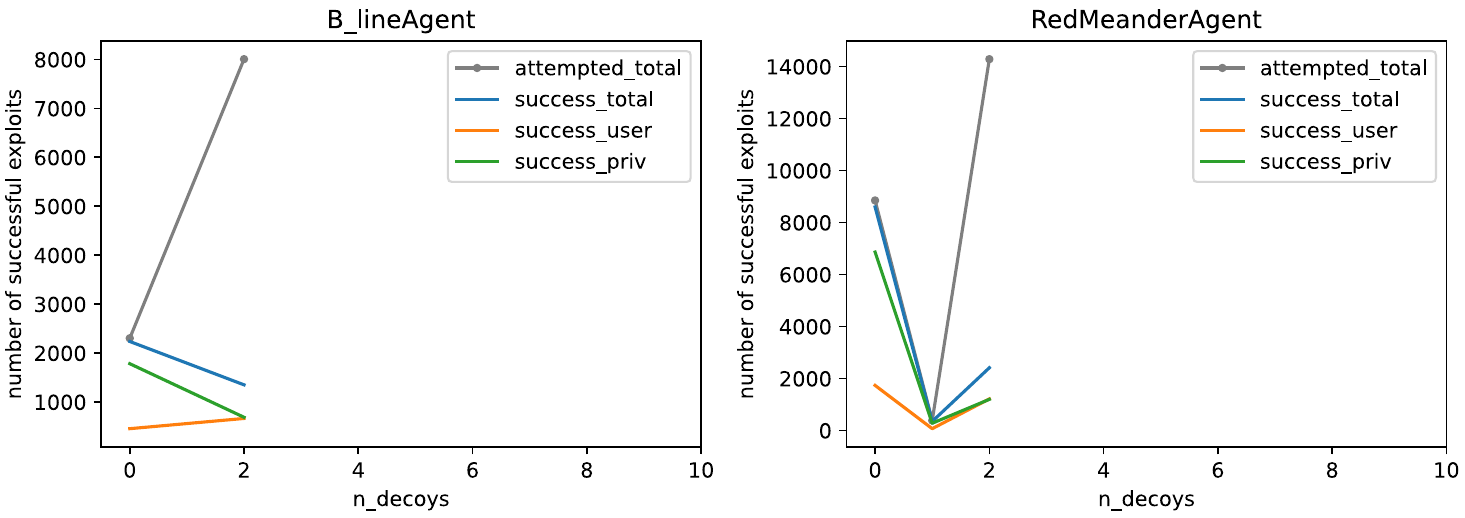}
     \vspace{-0.1in}
    \caption{(Mindrake, left; Cardiffuni, right):  Number of successful exploits on the User subnet vs number of decoys on the exploited host against each red agent.}
    \label{fig:ndecoys-user}
    
    \includegraphics[width=\decoyfigwidth]{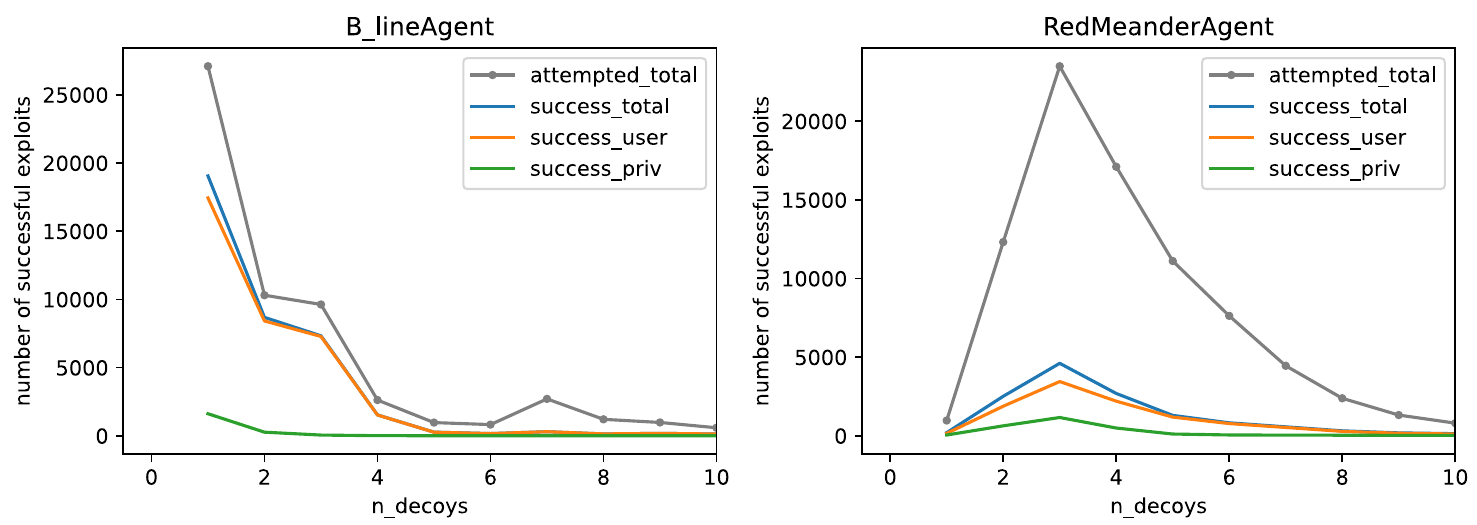}
    \hspace{1cm}
    \includegraphics[width=\decoyfigwidth]{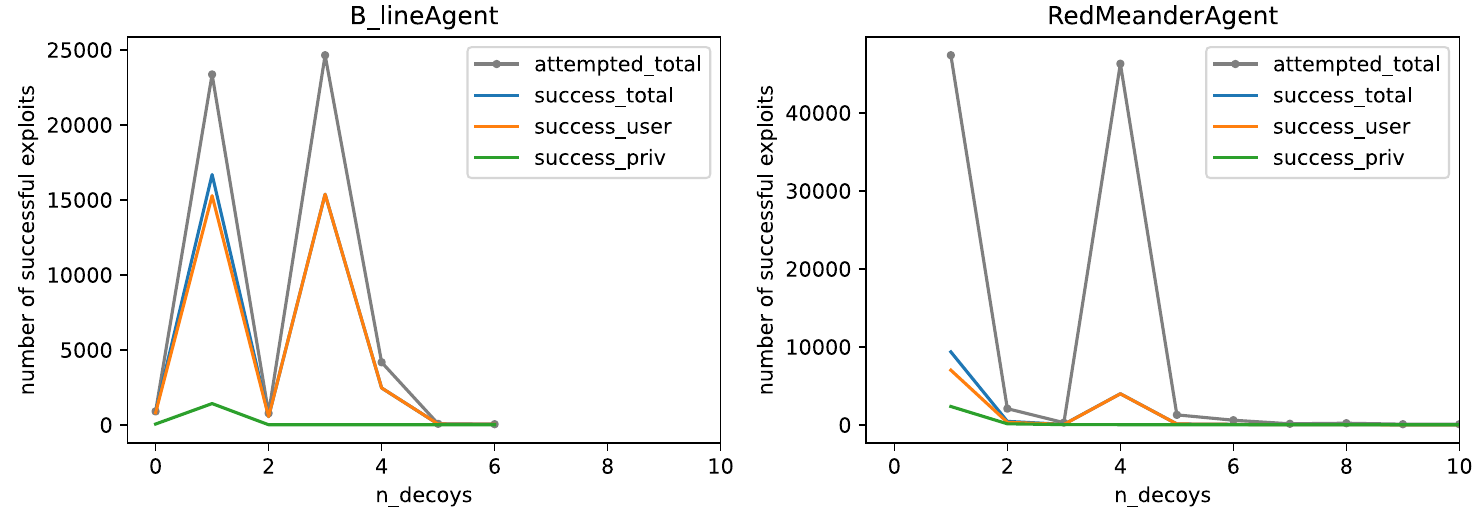}
     \vspace{-0.1in}
    \caption{(Mindrake, left; Cardiffuni, right):  Number of successful exploits on the Enterprise subnet vs number of decoys on the exploited host against each red agent.}
    \label{fig:ndecoys-enterprise}
\end{figure*}

\begin{figure}
    \centering
    \includegraphics[width=\decoyfigwidth, trim={0 0 0 1cm}, clip]{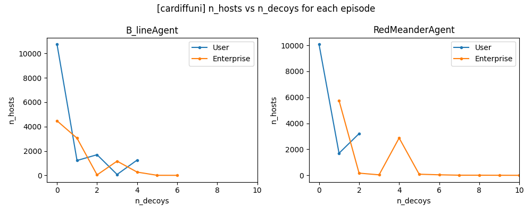}
     \vspace{-0.15in}
    \caption{
        Cardiffuni: Number of hosts on each subnet with the given number of decoys at the end of each episode.
        The peaks and valleys in these plots
        (particularly for n\_decoys\hspace{0.1em}$>$\hspace{0.1em}0)
        roughly correspond with the peaks and valleys in Figures~\ref{fig:ndecoys-user} ~(right)~and~\ref{fig:ndecoys-enterprise}~(right).
    }
    \label{fig:cardiffuni-nhosts-ndecoys}
\end{figure}

\section{Related Work}
Vyas et al.~\cite{vyas2023automated} present a review on automated cyber defense (ACD). In particular, the paper defines the key concepts in automated cyber defense, identifies the requirements for ACD agents and autonomous cyber operations gyms, examines the existing ACD work in the context of these requirements, and suggests research challenges for ACD. 
Autonomous cyber operation gyms, such as \cyberbattlesim~\cite{CyberBattleSim}, \cyborg~\cite{cyborg}, \farland~\cite{Farland-2021}, and \eireland~\cite{eireland-acd-2023}, are simulated and/or emulated network environments that facilitate the development and evaluation of ACD blue and red agents.

To further research in ACD, the TTCP CAGE working group has organized a series of challenge problems~\cite{cage_challenge}.
Foley et al.~\cite{foley2022automated} describe CAGE Challenge 1 (\cageone) and their Hierarchical PPO approach for blue agent design. Their system, called Mindrake, was the winner for~\cageone.
Wolk et al.~\cite{wolk2022cage} describe their approaches including Hierarchical PPO, attacker randomization, and ensemble RL for~\cagetwo, and examine their performance for unseen networks and unknown attacker strategies.
Kiely et al.~\cite{kiely2023autonomous} present a performance analysis for \cagetwo submissions, and categorize them based on their approaches: Single-Agent Deep Reinforcement Learning (DRL), Hierarchical DRL, Ensembles, and Non-DRL approaches. This paper is closely related to ~\cite{kiely2023autonomous}: this paper performs more in-depth analyses (e.g., state transition analysis) for the top-two submissions, while ~\cite{kiely2023autonomous} investigates the efficacy of the approaches used by all \cagetwo submissions.

Open-source evaluation platforms and public datasets such as the TTCP CAGE Challenges~\cite{cage_challenge} and
the DARPA Intrusion Detection System Evaluation~\cite{Lippmann2000EvaluatingID} can greatly
advance the science of cybersecurity by providing a common frame of reference to objectively compare the efficacy
of different approaches.
However, issues in the design of evaluation setups and procedures may weaken the outcome, as illustrated by
McHugh's critique of the DARPA IDS evaluation dataset~\cite{McHugh-IDSdataset-2000}.
To help ensure that the evaluation results may reflect the performance of the proposed solutions
in the real world, the realism of the challenge scenarios and model characteristics is important.
Vyas et al.~also independently noted the importance of the
realism of Autonomous Cyber Operation (ACO) Gyms and of deception techniques in their paper~\cite{vyas2023automated}.
The following section discusses the model realism aspect of CAGE challenges~(\cagetwo and \cagefour).

\section{Discussion: Model Realism and Future Work} \label{sec:discuss}

As one of the first challenge problems for studying the use of RL for autonomous network defense, CAGE Challenge 2 (\cagetwo) made some abstractions and simplifying assumptions to make the challenge more tractable. In this section, we identify several areas where there are gaps between the \cagetwo model and the real-world network defense environments. By explicitly identifying gaps, we may better understand the limitations of the proposed solutions and results, and suggest directions for future work. At the time of writing, CAGE Challenge 4 (\cagefour)~\cite{cage_challenge_4} has just started. We will highlight the differences between \cagetwo and \cagefour, where some of these gaps have been (partially) addressed. We first examine \cagetwo:
\begin{itemize}
    \item Scalability and host diversity: The network contains a small number of hosts and services.
    There is very limited diversity for hosts and services.

    \item Number and duration of actions: The red and blue agents are constrained to take one action for the entire network at each timestep.

    \item Blue agents: 
        The set of defense actions is limited. In particular,
        there are no actions to block attacker IPs, %
        or to redirect attacker connections to a honeypot (fishbowl), where one may observe the actions of the attacker.
        Repeated exploit attempts on an already accessed system do not seem to alert the blue agent.
   
    \item Green agents: 
        Blue agent actions (e.g., Remove and Restore) may negatively affect green agents.
        In~\cagetwo, there are fixed costs associated with such actions. For a more accurate model, the costs of those actions should depend on the numbers and types of green agents that are affected, particularly the cases that a remove or restore action is performed on a server that is used by green agents.

    \item Red agents: %
        When an attack against a service fails, the red agent does not learn from that result and tends to repeat the same attack against the same service.

    \item Decoy services: 
        Instead of deploying decoy services on production systems, it is common practice to use dedicated hosts to run decoy services to provide separation, so that attacks against decoys may not affect production systems. The costs for running decoy services (e.g., administrative and runtime costs) should be considered. Moreover, one may want to consider the tradeoffs between the realism of decoys (e.g., whether they may be fingerprintable) and the corresponding costs and risks.

    \item Operating environments:
        In \cagetwo, network connectivity, hosts, services, and the criticality of hosts/services are fixed.
        Moreover, \cagetwo assumes that the set of attacks are known and effective detection methods exist.
        In a realistic setting, systems and (attack and benign) activities may change over time.
        For novel attacks and new benign activities, the detection system may need to be adapted (e.g., using new features and signatures).

\end{itemize}

In \cagefour, some of the aforementioned limitations have been partially addressed.
\begin{itemize}
    \item Scalability and host diversity: Instead of having a single network, four networks are considered. These networks are more diverse, and of three different types, namely, (two) deployed networks, a headquarters network, and a contractor network. Moreover, each deployed network is divided into two security zones, the headquarters network into three zones, and the contractor network into one zone. For each zone, the number of hosts and their services will be randomized, and a zone has 1-6 servers and 3-10 user hosts. Also, each server will provide 1-5 services. 

    \item Number and duration of actions: Multiple blue agents are deployed, one for each security zone in deployed networks, and one for the headquarters network. There is a maximum of one red agent in each security zone, and each can maintain presence on multiple hosts. Actions will take different number of time steps, depending on their complexity. For example, the restore action for blue agents will take 5 time steps, and stealth service discovery action for red agents will take 3 time steps.

    \item Blue agents: \cagefour has added a commonly used defense action for blocking network traffic. Blue agents may deploy block or unblock traffic actions at the security zone level.

    \item Green agents: Behavior of green agents may trigger false alarms. If blue agent actions prevent green agents from doing their jobs, the defense will get a penalty.

    \item Red agents: To model commonly used adversarial tactics, \cagefour has added actions to perform stealth service discovery, to discover if a service is a decoy, to remove presence from a host, and to degrade services on a host if the red agent has root privileges on it.

    \item Decoy services: When a red agent tries to attack a decoy service in a security zone, the blue agent will detect and respond to it. 
    Unlike in \cagetwo, the blue agents in \cagefour have the ability to respond by blocking connections from the attacker's IP address at the security zone level.
    Also, red agents may use a DiscoverDeception action to detect decoy services.
    
    \item Operating environments: The number of hosts and services for the security zones are randomly generated. \cagefour uses a multi-phase mission scenario, where each phase has its own network access policy, and a different set of mission-critical hosts.
    
\end{itemize}

\section{Conclusions}

This paper analyzed two open-source DRL agents that participated in the CAGE Challenge 2. These agents were tasked with defending a simulated network against various rule-based attacks. By simplifying the complex state and action spaces, we gained valuable insights into the agents' behavior. Some of our key findings include the effectiveness of certain actions, the impact of decoy services, and the general ability of defenders to quickly clear infiltrations. We also discussed the realism of the CAGE Challenge 2 and the improvements made in CAGE Challenge 4 that partially address some of the realism concerns.

\scriptsize
\bibliographystyle{abbrv}
\bibliography{mainbib}

\end{document}